\documentclass{optica-article}

\journal{opticajournal} % for journals or Optica Open

\articletype{Research Article}

\usepackage{lineno}
\begin{document}

\title{Multi-mode Coherent Detection Ghost Imaging Lidar and Vibration-Mode Imaging}

\author{Jinquan Qi\authormark{1,2}, Shuang Liu\authormark{1}, Chenjin Deng\authormark{1,*}, Chaoran Wang\authormark{1,2}, Zunwang Bo\authormark{1}, Youzhen Gui\authormark{1,2}, Shensheng Han\authormark{1,2,3,**}}

\address{\authormark{1}Wangzhijiang Innovation Center for Laser, Aerospace Laser Technology and System Department, Shanghai Institute of Optics and Fine Mechanics, Chinese Academy of Sciences, Shanghai 201800, China}

\address{\authormark{2}Center of Materials Science and Optoelectronics Engineering, University of Chinese Academy of Sciences, Beijing 100049, China}

\address{\authormark{3}Hangzhou Institute for Advanced Study, University of Chinese Academy of Sciences, Hangzhou 310024, Zhejiang, China}

\email{\authormark{*}dcj@siom.ac.cn} %% email address is required; see note below about the corresponding author designation
\email{\authormark{**}sshan@mail.shcnc.ac.cn}
% use {asbstract*} to suppress the copyright line. Copyright information will be added in production

\begin{abstract*} 
Coherent detection ghost imaging lidar (CD-GI lidar) integrates ghost imaging with coherent detection, thereby achieving enhanced anti-interference and phase-resolved imaging capability. Here, we propose a bucket-detector-based multi-mode coherent detection scheme for CD-GI lidar, where the reflected multi-mode light fields are coherently mixed with a single-mode local oscillator (LO) at the bucket detector photosensitive plane. The bucket-detector-based multi-mode CD-GI lidar system breaks the constraints of Siegman antenna theorem by utilizing field correlation to decouple the reflected multi-mode light fields and reconstructs the spatial distribution of targets' vibration modes. Theoretical analysis of the bucket-detector-based multi-mode CD-GI lidar system is presented in this work, and its feasibility is verified through a series of experiments.
\end{abstract*}

%%%%%%%%%%%%%%%%%%%%%%%%%%  body  %%%%%%%%%%%%%%%%%%%%%%%%%%
\section{Introduction}
Ghost imaging\cite{shapiro2012physics,qiu2023remote,zeng2025tailoring} has attracted significant interest in recent decades, as it can employ optical field fluctuation to encode and compress high-dimensional spatial image information into a lower-dimensional detection space that can be directly measurable by a detector, and allows the detection and imaging processes to be physically separated. Ghost imaging separates the tasks of optical signal acquisition and spatial resolution, enabling spatially resolved imaging performance comparable to that of detector arrays while operating within the spatiotemporal bandwidth product of a single detector. These advantages make it suitable for non-traditional imaging scenarios, including staring imaging radar and lidar\cite{zhao2012ghost,erkmen2012computational,2013Computational,cheng2016radar,wang2018nonrandom,sun2025quantum}, ultra-low dose X-ray imaging\cite{yu2016}, electron\cite{li2018electron}, and neutron radiographic imaging\cite{kingston2020neutron,he2021single}, as well as atomic ghost imaging
\cite{khakimov2016ghost}, snapshot multidimensional ghost imaging\cite{liu2016spectral,wang2022dispersion}, super-resolution imaging based on spectral dimensional information\cite{chen2025multicolor}, and scattering imaging\cite{2010Correlated,2011Reflective,gong2011correlated,bina2013backscattering,li2023underwater,zhang2023different}, etc. While implementations based on intensity-correlations only can reconstruct the target’s intensity image, subsequent developments in field correlation ghost imaging employing homodyne detection have shown that both quantum-entangled light sources\cite{2004Ghost} and classical thermal light sources\cite{2009Homodyne} are capable of retrieving the complex reflectivity of the target.

Recent advancements in optical coherent detection\cite{ip2008coherent,taylor2009phase,kikuchi2015fundamentals,fink1975coherent} have enabled sensitivities approaching the shot-noise limit, comparable to intensity detection, while simultaneously offering the capability to suppress background light interference and retrieve both the amplitude and phase information of the light field. These capabilities have motivated the development of coherent detection imaging lidar systems, notably: (i) synthetic aperture imaging lidar (SAIL), which circumvents the diffraction limit to achieve high spatial resolution through synthetic aperture synthesis\cite{beck2005synthetic,liu2006coherent,pellizzari2009synthetic,barber2014synthetic,shi2024motion}, and (ii) coherent focal-plane array (FPA) imaging lidar, which enables real-time parallel detection with high detection efficiency\cite{Wickham2000,Hemmat2010,McManamon2012,Yariv2014,Katz2017,Xie2020,Zhu2022}. However, the fundamental limitation of SAIL is that it relies on platform-induced relative motion to modulate the return signal. Concurrently, coherent FPA imaging is challenged by the spatiotemporal bandwidth product required for both large-scale detection and high-speed data processing\cite{simpson1997coherent,chen2015fpa}.

The integration of ghost imaging and coherent detection provides a promising alternative to the aforementioned coherent-detection–based imaging lidar systems\cite{deng2016pulse,pan2020experimental}, which can retrieve spatial resolution via field correlation using only a single-bucket detector, alleviate the spatiotemporal bandwidth requirements on the receiver, and enable staring imaging without platform-induced relative motion. In this work, through theoretical analysis and experiments, we show that the bucket-detector-based multi-mode CD-GI lidar system overcomes the limitation imposed by the Siegman antenna theorem through field correlation, decoupling the coherent summation of reflected multi-mode light fields. It enables the simultaneous retrieval of both amplitude and phase information of the target, and provides access to the spatial distribution of targets' vibration modes in dynamic scenes within the field of view (FoV). Ultimately, our approach offers a framework for staring imaging with reduced spatiotemporal bandwidth product requirements.

\section{Method and Analysis} \label{Theoretical}

\begin{figure}[htbp]
    \centering
    \includegraphics[width=\linewidth]{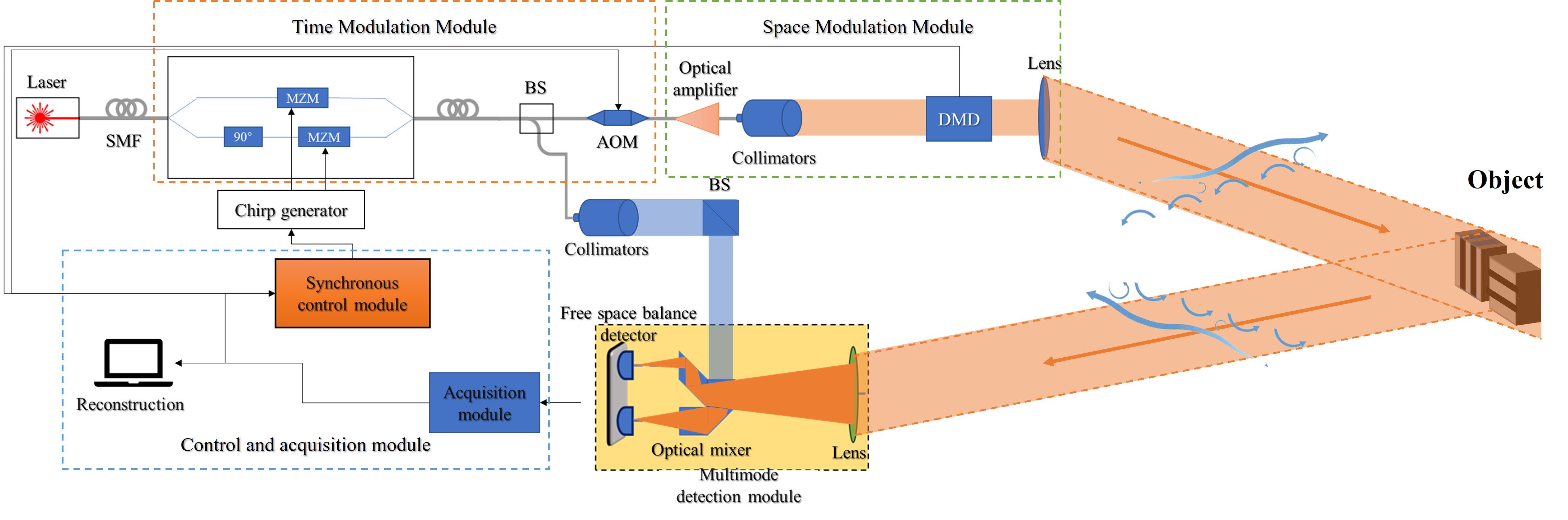}
    \caption{Schematic of the bucket-detector-based multi-mode CD-GI lidar system.}
    \label{fig:system}
\end{figure}
Fig.~\ref{fig:system} illustrates the bucket-detector-based multi-mode CD-GI lidar system. A spatiotemporally modulated laser field illuminates the target, and the resulting reflected multi-mode light fields are collected by a bucket detector and coherently mixed with a LO to produce an intermediate-frequency (IF) signal containing the target’s amplitude and phase information.

\subsection{Light field transmission and coherent detection}\label{Derivation for Vibrating Targets}
The spatiotemporally modulated light field is generated by the coordinated temporal and spatial modulation modules illustrated in Fig.~\ref{fig:system}. A narrow-linewidth laser serves as the seed source, yielding an emitted light field described by $E_0 = A_0 \exp{(j2 \pi f_c t)}$, where $f_c$ denotes the center frequency. To achieve distance resolution capability, the seed light is initially modulated by an electro-optic modulator (EOM) driven by a chirped radio-frequency signal $s(t)$. The resultant beam is partitioned by a fiber beam splitter into an object arm and a LO arm. In the object arm, an acousto-optic modulator (AOM) induces a frequency shift $f_{if}$. Subsequently, the temporally modulated beam is collimated and incident onto a digital micromirror device (DMD) operating in an externally triggered mode, synchronized with the temporal modulation period. The DMD is intrinsically a binary amplitude modulator, and generates a series of predefined spatial encoding patterns denoted as $E_s^{(i)}(\boldsymbol{\rho}_s) = A_s^{(i)}\left( \boldsymbol{\rho}_s\right) \exp\left[ j \phi_s^{(i)}\left( \boldsymbol{\rho}_s\right) \right]$. Since the DMD is a pre-programmed encoding device acting only on the object arm, the reference-arm field can be defined through a virtual encoding formulated as: $E_r^{(i)}(\boldsymbol{\rho}_r ) = A_s^{(i)}\left( \boldsymbol{\rho}_r\right) \exp\left[ j \phi_s^{(i)}\left( \boldsymbol{\rho}_r\right) \right]$. 
This architecture produces a spatiotemporally modulated transmitted field, which can be expressed as:
\begin{equation}
E_\text{Sig}^{(i)}(\boldsymbol{\rho}_s,t) = \sqrt{\eta_\mathrm{Sig}} A_0 \cdot \exp\left[j2 \pi (f_c+f_{if}) t\right] \cdot s(t) \cdot E_s^{(i)}(\boldsymbol{\rho}_s),
\label{eq:Ed_spacetempoal}
\end{equation}
and a LO field:
\begin{equation}
E_{\text{Lo-mod}}(t) = \sqrt{\eta_\mathrm{Lo}}\,A_0 \cdot \exp\left(j2 \pi f_c t\right) \cdot s(t),
\label{eq:lo1}
\end{equation}
where $i$ denotes the $i$-th pulse, $\eta_\mathrm{Sig}\approx 0.99$ and $\eta_\mathrm{Lo}\approx 0.01$ denote the power splitting ratios.

After propagation through the imaging transmission module, as shown in Fig.~\ref{fig:system}, the modulated field illuminates the target, which is modeled by a spatially varying complex reflectivity combined with a time-dependent phase term accounting for surface vibrations:
\begin{equation}
    \tilde{T}(\boldsymbol{\rho}_o,t) = t_o(\boldsymbol{\rho}_o) \exp\left[ j \phi_o(\boldsymbol{\rho}_o) \right]\cdot\exp\left[ j 2 k z(\boldsymbol{\rho}_o,t) \right] ,
    \label{eq:lo}
\end{equation}
where \( t_o(\boldsymbol{\rho}_o) \) indicates the amplitude reflectivity of target, \( \phi_o(\boldsymbol{\rho}_o) \) represents the phase of target and \( z(\boldsymbol{\rho}_o,t) \) indicates the distance variation caused by the micro-vibration of target.
The resulting multi-mode reflected light field is collected by the imaging receiving module and coherently interfered with the LO in a spatial optical mixer, followed by the balanced bucket detector. The architecture of this multi-mode detection module is described in Fig.~\ref{fig:system}. The reflected light fields on the surface of the detector can be represented as (see Supplement 1, Sec.~S1 for detailed derivation process):
\begin{align}
E_d^{(i)}&(\boldsymbol{\rho}_d,t) =\  
\frac{\exp\left[jk(z_1 + z_2 + z_3 + z_4)\right]}{ \lambda^4 z_1 z_2 z_3 z_4} \exp\left[ \frac{j\pi}{\lambda z_4} |\boldsymbol{\rho}_d|^2 \right] s(t - \tau) \cdot \exp[j(\phi_{i,1}+\phi_{i,2})] \notag \\
&\times \int_{A_o} 
\exp\left[
\frac{j \pi}{\lambda} \left( \frac{z_1 + z_2}{z_2^2} + \frac{1}{z_3} \right) |\boldsymbol{\rho}_o|^2 
\right] E_s^{(i)}\left(-\frac{z_1}{z_2} \boldsymbol{\rho}_o \right) 
\tilde{T}(\boldsymbol{\rho}_o,t)  \frac{J_1 \left( \frac{D_f \pi}{\lambda} 
\left( \frac{\boldsymbol{\rho}_o}{z_3} + \frac{\boldsymbol{\rho}_d}{z_4} \right) \right)}
{\frac{\pi}{\lambda} \left( \frac{\boldsymbol{\rho}_o}{z_3} + \frac{\boldsymbol{\rho}_d}{z_4} \right)} 
\, 
d\boldsymbol{\rho}_o 
\label{eq:sig}
\end{align}
where $k = 2\pi / \lambda$ designates the wave number; $z_1$ and $z_2$ represent the object and image distances of the imaging transmission module, while $z_3$ and $z_4$ refer to the corresponding distances for the imaging receiving module. The term $\exp[j(\phi_{i,1}+\phi_{i,2})]$ accounts for the overall phase distortion introduced by atmospheric turbulence during the propagation of the $i$-th spatially modulated light field, with $\phi_{i,1}$ and $\phi_{i,2}$ characterizing the forward and return paths, respectively. Additionally, $D_f$ denotes the aperture diameter of the receiving lens, and $A_o$ signifies the target area. Finally, $\boldsymbol{\rho}_o=(x_o,y_o)$ stands for the transverse coordinates on the target plane, whereas $\boldsymbol{\rho}_{d}=(x_{d},y_{d})$ indicates the coordinates on the detector plane. After balanced detection, the optical signal is converted into an electrical signal $i^{(i)}(t)$:
\begin{equation}
    \begin{aligned}
    i^{(i)}(t) &= \frac{1}{2} R \cdot \int_{A_d} \left[ E_d^{(i)}(\boldsymbol{\rho}_d,t) E_{\text{Lo-mod}}^*(t) + c.c. \right]d\boldsymbol{\rho}_d \\
    &= R \cdot \int_{A_d} \int_{A_o} A_1(\boldsymbol{\rho}_o, \boldsymbol{\rho}_d) \cos\left[ \phi(\boldsymbol{\rho}_o, \boldsymbol{\rho}_d,t) \right] d\boldsymbol{\rho}_o d\boldsymbol{\rho}_d.
    \label{eq:dianxinhao}
    \end{aligned}
\end{equation}
\begin{align*}
    &A_1(\boldsymbol{\rho}_o, \boldsymbol{\rho}_d) =\frac{\sqrt{\eta_\mathrm{Sig}} \sqrt{\eta_\mathrm{Lo}}}{\lambda^4 z_1  z_2  z_3  z_4} A_s^{(i)}\left(-\frac{z_1}{z_2} \boldsymbol{\rho}_o\right) A_0^2 t_o(\boldsymbol{\rho}_o) \cdot \frac{ J_1\left( \frac{D_f \pi}{\lambda} \left( \frac{\boldsymbol{\rho}_o}{z_3} + \frac{\boldsymbol{\rho}_d}{z_4} \right) \right)}{\frac{\pi}{\lambda} \left( \frac{\boldsymbol{\rho}_o}{z_3} + \frac{\boldsymbol{\rho}_d}{z_4} \right)} && \tag{\ref{eq:dianxinhao}{.1}}\\
    &\phi(\boldsymbol{\rho}_o, \boldsymbol{\rho}_d ,t) =k(z_1 + z_2 + z_3 + z_4)+ \frac{\pi}{\lambda z_4} |\boldsymbol{\rho}_d|^2  +\phi_o(\boldsymbol{\rho}_o) + \phi_s^{(i)}\left(-\frac{z_1}{z_2} \boldsymbol{\rho}_o\right) + 2\pi f_{if} t\\
    & + \frac{\pi}{\lambda} \left( \frac{z_1 + z_2}{z_2^2} + \frac{1}{z_3} \right) |\boldsymbol{\rho}_o|^2 + 2\pi\left(\frac{B}{T} \tau t + f_0 \tau - \frac{B}{2T} \tau^2 \right) + 2kz(\boldsymbol{\rho}_o,t) + \phi_{i,1}+\phi_{i,2} && \tag{\ref{eq:dianxinhao}{.2}}
\end{align*}
Here, \( \tau = (z_1+z_2+z_3+z_4)/{c} \) is the time delay due to the propagation of light; \( R \) denotes the responsivity of the detector; \( c.c. \) denotes the complex conjugate of $E_d^{(i)}(\boldsymbol{\rho}_d,t) E_{\text{Lo-mod}}^*(t)$ and \( A_d \) represents the area of the detector.

\subsection{Signal processing and Vibration-Mode imaging reconstruction} \label{signal_process}
In practical remote-sensing scenarios, many targets exhibit micro-vibration motions, such as engine-induced vibrations\cite{barton2024noise}, rotor oscillations\cite{simonovskiy2021methods}, and structural flutter\cite{chai2021aeroelastic}, which arise from continuous external excitation.

Without loss of generality, an arbitrary vibration can be decomposed into a superposition of sinusoidal components\cite{feldman2011hilbert}. Consequently, we adopt a canonical sinusoidal vibration model along the axial direction:
\begin{equation}
z(\boldsymbol{\rho}_o,t) = \Delta z(\boldsymbol{\rho}_o) \sin\left[2\pi f_v t + \varphi_o(\boldsymbol{\rho}_o)\right],
\end{equation}
where $\Delta z(\boldsymbol{\rho}_o)$ denotes the vibration height, $f_v$ is the vibration frequency and $\varphi_o(\boldsymbol{\rho}_o)$ represents the initial vibration phase.

To simplify the problem, the drive frequency of target is set as an integer multiple of the DMD repetition rate, ensuring synchronization with the target vibration. The detected signal $i^{(i)}(t)$ is then subjected to a Short-Time Fourier Transform (STFT). The \( i^{(i)}(t) \) is then subjected to a short-time Fourier transform (STFT). Assuming the total length of \( i^{(i)}(t) \) is $L$, a sliding rectangular window function $w(t)$ is used with length $l$ and overlap $d$, $k$ is the number of the corresponding windows: $k = 0, 1, 2, \dots, \left\lfloor \frac{L - l}{l - d} \right\rfloor$, the result of the STFT can be expressed as:
\begin{equation}
    \begin{aligned}
    \widetilde{i}_k^{(i)}\left(f = \frac{\omega_k^{(h)}}{2 \pi} \right) &= Rl \int_{A_d}\int_{A_o} A_1(\boldsymbol{\rho}_o, \boldsymbol{\rho}_d) \pi e^{j \Phi(\boldsymbol{\rho}_o, \boldsymbol{\rho}_d)} e^{-j \omega_k^{(h)} \left[k(l - d) + \frac{l}{2} \right]} \mathrm{sinc}(\frac{\omega_k^{(h)} l}{2})  d\boldsymbol{\rho}_o d\boldsymbol{\rho}_d \\
       &+Rl \int_{A_d} \int_{A_o} A_1(\boldsymbol{\rho}_o, \boldsymbol{\rho}_d) \pi e^{-j \Phi(\boldsymbol{\rho}_o, \boldsymbol{\rho}_d)} e^{j \omega_k^{(h)} \left[k(l - d) + \frac{l}{2} \right]}  \mathrm{sinc}(\frac{\omega_k^{(h)} l}{2}) d\boldsymbol{\rho}_o d\boldsymbol{\rho}_d,
       \label{eq:stftr}
    \end{aligned}
\end{equation}
\begin{align*}
    &A_1(\boldsymbol{\rho}_o, \boldsymbol{\rho}_d) =\frac{\sqrt{\eta_\mathrm{Sig}} \sqrt{\eta_\mathrm{Lo}}}{\lambda^4 z_1  z_2  z_3  z_4} A_s^{(i)}\left(-\frac{z_1}{z_2} \boldsymbol{\rho}_o\right) A_0^2 t_o(\boldsymbol{\rho}_o) \cdot \frac{ J_1 \left( \frac{D_f \pi}{\lambda} \left( \frac{\boldsymbol{\rho}_o}{z_3} + \frac{\boldsymbol{\rho}_d}{z_4} \right) \right)}{\frac{\pi}{\lambda} \left( \frac{\boldsymbol{\rho}_o}{z_3} + \frac{\boldsymbol{\rho}_d}{z_4} \right)}  \tag{\ref{eq:stftr}{.1}}
\end{align*}
\begin{align*}
    \Phi(\boldsymbol{\rho}_o, \boldsymbol{\rho}_d) &= k(z_1 + z_2 + z_3 + z_4) + \phi_o(\boldsymbol{\rho}_o) + \phi_s^{(i)}\left(-\frac{z_1}{z_2} \boldsymbol{\rho}_o \right) + \frac{\pi}{\lambda z_4} |\boldsymbol{\rho}_d|^2 \\
    &+ \frac{\pi}{\lambda} \left( \frac{z_1 + z_2}{z_2^2} + \frac{1}{z_3} \right) |\boldsymbol{\rho}_o|^2 + 2\pi\left( f_0 \tau - \frac{B}{2T} \tau^2 \right) + 2k z_0(\boldsymbol{\rho}_o) + \phi_i \tag{\ref{eq:stftr}{.2}}
\end{align*}
where \( \omega_k^{(h)} = 2\pi \left[ f_{if} + \frac{B}{T} \tau + \frac{4\pi f_v \Delta z(\boldsymbol{\rho}_o)}{\lambda} \cos \left[ 2\pi f_v \cdot k(l - d) + \phi_o(\boldsymbol{\rho}_o) \right] \right] \) is the corresponding to the $h$-th micro-Doppler frequency in the $k$-th time slice of STFT.

Atmospheric turbulence correction has been a critical focus in coherent detection techniques, with many previous works \cite{chen2018fso,martinez2024self,Hersbach2020ERA5,Hu2019_atmospheric_delay_DInSAR_ERA5} addressing this issue. In this work, we also consider turbulence correction using a time-division strategy\cite{Zhang2021TurbulencePilot}, where even frames correct the odd frames in the encoding process. The detailed derivation is provided in Supplement 1, Sec.~S2.A. The corrected IF signal expression is given as follows:
\begin{equation}
\begin{aligned}
    \widetilde{i}_{\mathrm{correct}-k}^{(i)} \left(f = \frac{\omega_k^{(h)}}{2 \pi} \right) = \frac{\widetilde{i}_k^{(2i-1)} \left(f = \frac{\omega_k^{(h)}}{2 \pi} \right)}{\exp{\left[j \cdot \mathrm{arg} (\widetilde{i}_k^{(2i)}  \left(f = \frac{\omega_k^{(h)}}{2 \pi} \right))\right]}}
\label{correct_signal}
\end{aligned}
\end{equation}
After performing field correlation of the corrected IF signal
$\widetilde{i}_{\mathrm{correct}-k}^{(i)}\!\left(f = \omega_k^{(h)}/2\pi\right)$ with the reference-arm field $\left[ E_r^{(i)}(\boldsymbol{\rho}_r) \right]^{*}$, we can obtain the spatial distribution corresponding to the micro-Doppler frequency \(f = \omega_k^{(h)} / 2 \pi \) of the target:
\begin{equation}
\begin{aligned}
    G_k^{(h)}&(\boldsymbol{\rho}_r) = {\left\langle \widetilde{i}_{\mathrm{correct}-k}^{(i)} \left(f = \frac{\omega_k^{(h)}}{2 \pi} \right) \cdot \left[ E_r^{(i)}(\boldsymbol{\rho}_r) \right]^{*} \right\rangle}_i \\
    &\propto A_0^2 \pi l R \frac{\sqrt{\eta_\mathrm{Sig}} \sqrt{\eta_\mathrm{Lo}}}{\lambda^4 z_1  z_2  z_3  z_4}  \exp\left[ jk(z_1 + z_2 + z_3 + z_4)-j\omega_k^{(h)} \left[k(l - d) + \frac{l}{2} \right] \right] \mathrm{sinc}(\frac{\omega_k^{(h)} l}{2})  \\
    &\times \exp\left[j 2\pi\left( f_0 \tau - \frac{B}{2T} \tau^2 \right) \right]  \exp\left[ \frac{j \pi}{\lambda} \left( \frac{z_1 + z_2}{z_2^2} + \frac{1}{z_3} \right) |-\frac{z_2}{z_1}\boldsymbol{\rho}_r|^2 \right] \frac{ J_1 \left[\frac{D_f \pi}{\lambda} \left( \frac{z_2}{z_1z_3}\boldsymbol{\rho}_r \right) \right]}{\frac{\pi}{\lambda} \left( \frac{z_2}{z_1z_3}\boldsymbol{\rho}_r \right)} \\
    &\times \tilde{T}_k^{(h)}(-\frac{z_2}{z_1}\boldsymbol{\rho}_r) .
    \label{eq:imaging}
\end{aligned}
\end{equation}
Here, the ensemble averaging \(\langle \cdot \rangle\) is performed over the encoded optical field $i$ corresponding to the $h$-th micro-Doppler frequency component within the $k$-th time window.
The vibration height corresponding to the target spatial position can be converted as follows (see Supplement 1, Sec.~S2.B for detailed derivation process):
\begin{equation}
\begin{aligned}
    A_{k}^{(h)}(\boldsymbol{\rho}_r) = \frac{\left[\frac{\omega_k^{(h)}}{2 \pi}-f_{if}-\frac{B}{T} \tau\right] \lambda}{4\pi  f_v}.
    \label{eq:ampli}
\end{aligned}
\end{equation}
The spatial images \( G_k^{(h)}(\boldsymbol{\rho}_r) \) reconstructed from these windows are combined with the corresponding vibration height \( A_{k}^{(h)}(\boldsymbol{\rho}_r) \) to recover the target’s vibration modes. 

A static target represents a special case where the vibration frequency vanishes ($f_v = 0$), yielding zero micro-Doppler shifts across all temporal slices. In this scenario, the STFT processing degenerates into a standard FFT, and the complex reflectivity reconstruction result in Eq.~\eqref{eq:imaging} can be simplified as: 
\begin{equation}
\begin{aligned}
    G(\boldsymbol{\rho}_r) &= \left\langle \widetilde{i}_{\mathrm{correct}}^{(i)} \left(f = f_{if} + \frac{B}{T}\,\tau\right) \cdot \left[ E_r^{(i)}(\boldsymbol{\rho}_r) \right]^* \right\rangle \\
    &\propto A_0^2 \pi l R \frac{\sqrt{\eta_\mathrm{Sig}} \sqrt{\eta_\mathrm{Lo}}}{\lambda^4 z_1  z_2  z_3  z_4}  \exp\left[ jk(z_1 + z_2 + z_3 + z_4) \right] \exp\left[j 2\pi\left( f_0 \tau - \frac{B}{2T} \tau^2 \right) \right]  \\
    &\times   \exp\left[ \frac{j \pi}{\lambda} \left( \frac{z_1 + z_2}{z_2^2} + \frac{1}{z_3} \right) |-\frac{z_2}{z_1}\boldsymbol{\rho}_r|^2 \right] \tilde{T}(-\frac{z_2}{z_1}\boldsymbol{\rho}_r) \frac{ J_1 \left[\frac{D_f \pi}{\lambda} \left( \frac{z_2}{z_1z_3}\boldsymbol{\rho}_r \right) \right]}{\frac{\pi}{\lambda} \left( \frac{z_2}{z_1z_3}\boldsymbol{\rho}_r \right)} .
    \label{eq:17}
\end{aligned}
\end{equation}
The derivation of the field correlation reconstruction for static targets is provided in Supplement 1, Sec.~S2.A.

\subsection{Analysis}
For the imaging receiving module, the maximum angular FoV is governed by $\theta_{\mathrm{FoV}} \approx D_{\mathrm{det}} / f$, where $D_{\mathrm{det}}$ designates the effective detector width and $f$ represents the focal length of the receiving lens. Consequently, the linear FoV on the target plane scales proportionally with the detector size and can be formulated as $\mathrm{FoV} \approx \theta_{\mathrm{FoV}} \cdot z_3 \approx D_{\mathrm{det}} \cdot M$, where $M = z_3 / z_4$ characterizes the magnification of the receiving module. It follows that increasing the detector size directly expands the accessible imaging region on the target plane.

\subsubsection{Analysis for the IF signal energy}
The IF signal can be reformulated in the wave-vector domain by expressing both the LO fields and the reflected multi-mode light fields as described in Eqs.\eqref{eq:lo1} and \eqref{eq:sig} using angular spectrum representation\cite{Zhong2023ModifiedASM,Song2024FullVectorASM}. The detailed derivation is provided in Supplement~1, Sec.~S3.A. The resulting expression can be indicated as:
\begin{equation}
\begin{split}
    \widetilde{i}^{{(i)}}(f&=f_{if})=R \sqrt{\eta_\mathrm{Sig}} \sqrt{\eta_\mathrm{Lo}} A_0^2
        \frac{\exp\left[jk(z_1 + z_2 + z_3 + z_4)\right]}{ \lambda^4 z_1 z_2 z_3 z_4}  \\
        &\times \int_{|\boldsymbol{\kappa}|} \int_{A_o} 
        \exp\left[ \frac{j \pi}{\lambda} \left( \frac{z_1 + z_2}{z_2^2} + \frac{1}{z_3} \right) |\boldsymbol{\rho}_o|^2 \right] E_s^{{(i)}}\left(-\frac{z_1}{z_2} \boldsymbol{\rho}_o \right) \tilde{T}(\boldsymbol{\rho}_o)  \exp\left[ -j2 \pi \frac{z_4}{z_3} \boldsymbol{\rho}_o \boldsymbol{\kappa} \right] d\boldsymbol{\rho}_o  \\
        &\times  \prod(\frac{|\boldsymbol{\kappa}|}{\kappa_c}) \exp\left[ j\pi \lambda z_4 |\boldsymbol{\kappa}|^2 \right] \pi \left(\frac{D_{\mathrm{det}}}{2}\right)^2 \frac{2J_1 \left( \frac{D_{\mathrm{det}} }{2} |\boldsymbol{\kappa}-\boldsymbol{\kappa}_{\mathrm{lo}}| \right)}{\frac{D_{\mathrm{det}} }{2} |\boldsymbol{\kappa}-\boldsymbol{\kappa}_{\mathrm{lo}}|} d\boldsymbol{\kappa} ,
        \label{eq:f_if}
\end{split} 
\end{equation}
\begin{align}
\prod(\frac{|\boldsymbol{\kappa}|}{\kappa_c})=
\begin{cases}
1, & |\boldsymbol{\kappa}| \leq \kappa_c, \\[6pt]
0, & \text{otherwise}.
\end{cases} \qquad \kappa_c = \frac{D_f}{ \lambda z_4}  \tag{\ref{eq:f_if}{.1}}
\end{align}
where $\kappa_c = \frac{D_f}{ \lambda z_4}$ represents the cutoff spatial frequency of the imaging receiving module; $\boldsymbol{\kappa}_{\mathrm{lo}}$ denotes the transverse wave vector (angular spectrum component) of the single-mode LO; and $D_{\mathrm{det}}$ is the diameter of the bucket detector. This integral expression in terms of $\boldsymbol{\kappa}$ explicitly shows that the detector area exerts a spatial-frequency weighting on the IF signal, thereby dictating whether the detection process operates in the single-mode or the multi-mode regime.

\begin{enumerate}

\item \textbf{Single-mode light field:}  
When $D_{\mathrm{det}} \leq 2 \lambda z_4/D_f$, Eq.\eqref{eq:f_if} can be simplified as:
\begin{equation}
\begin{aligned}
    \widetilde{i}^{{(i)}}(f=f_{if}) \propto \int_0^{\kappa_c} \kappa\, \prod\left(\frac{\kappa}{\kappa_c}\right)  \pi \left(\frac{D_{\mathrm{det}}}{2}\right)^2 d\kappa .
    \label{eq:sighenb}
\end{aligned} 
\end{equation}
In this case, the detector diameter is smaller than the diffraction-limited Airy disk, enabling the detector to be approximated as a point detector capturing only a single-mode field. As the detector size increases, the energy of the IF signal scales accordingly as described in Eq.~\eqref{eq:sighenb}, which is consistent with the Siegman antenna theorem \cite{siegman1966antenna}.

\item \textbf{Multi-mode light field:}  
When $D_{\mathrm{det}} > 2 \lambda z_4/D_f$, Eq.\eqref{eq:f_if} is simplified as:
\begin{equation}
\begin{aligned}
    \widetilde{i}^{{(i)}}(f=f_{if}) \propto \int_0^{\frac{\pi}{2}} \int_0^{\kappa_c} \kappa\, \prod\left(\frac{\kappa}{\kappa_c}\right)  \frac{2J_1\!\big(\tfrac{D_{\mathrm{det}}}{2} \kappa \theta\big)} {\tfrac{D_{\mathrm{det}}}{2} \kappa \theta} d\kappa d\theta .
    \label{eq:sighenb1}
\end{aligned} 
\end{equation}
    where $\theta$ denotes the angle between the wave-vectors of the LO field and the reflected multi-mode light fields. As the detector diameter exceeds the Airy disk, it functions as a bucket detector that integrates multiple spatial modes. Consequently, the IF signal energy statistically saturates into a stable oscillatory regime, further validating the predictions of the Siegman antenna theorem \cite{siegman1966antenna}.
\end{enumerate}

% In summary, the coherent detection performance is governed by the spatial-mode matching between the reflected light fields and LO fields, together with the detector’s effective area, as described by Siegman antenna theorem. A small detector favors single-mode detection, whereas increasing the detector size enables the collection of multiple spatial modes. In the presence of reflected multi-mode light fields, the IF signal energy rises to saturation and then oscillates as described in Siegman antenna theorem.
In summary, the coherent detection performance is governed by the spatial-mode matching between the reflected light fields and the LO fields, contingent upon the detector’s effective size, as dictated by the Siegman antenna theorem. While a constrained detector size is conducive to single-mode detection, expanding the detector size facilitates the collection of multiple spatial modes. In the presence of spatially complex reflected multi-mode fields, the IF signal energy initially increases toward saturation and subsequently enters an oscillatory regime, a behavior consistent with the theoretical predictions of the Siegman antenna theorem.

\subsubsection{Analysis for the energy of reconstructed image by field correlation}
The field correlation can be formulated in the wave-vector domain as:
\begin{align}
G[\boldsymbol{\rho}_r^{(m,n)}]
& \propto
\sum_{m^{'},n^{'}}
\Bigg\{
\tilde{t}_o\!\left( \alpha\,\boldsymbol{\kappa}_{m^{'},n^{'}} \right)
*
\left[
\mathrm{sinc}\!\left( \beta\, \boldsymbol{\kappa}_{m^{'},n^{'}} \right)
e^{j 2\pi \gamma\, \boldsymbol{\rho}_r^{(m,n)} \!\cdot\!\boldsymbol{\kappa}_{m^{'},n^{'}}}
\right]
\Bigg\} \notag
\\
&\times  
\Delta\kappa^2\,
\prod\!\left(\frac{|\boldsymbol{\kappa}_{m^{'},n^{'}}|}{\kappa_c}\right)
\pi\left(\frac{D_{\mathrm{det}}}{2}\right)^2
\frac{
2 J_1\!\left( \tfrac{D_{\mathrm{det}}}{2} |\boldsymbol{\kappa}_{m^{'},n^{'}}-\boldsymbol{\kappa}_{\mathrm{lo}}| \right)
}{
\tfrac{D_{\mathrm{det}}}{2} |\boldsymbol{\kappa}_{m^{'},n^{'}}-\boldsymbol{\kappa}_{\mathrm{lo}}|
},
\label{eq:figuer}
\end{align}
\begin{align*}
\alpha = \frac{z_4}{z_3}, \beta = \frac{z_4}{z_3}\frac{z_2}{z_1}, \gamma =\frac{z_4}{z_3z_1}, \tilde{t}_o(\alpha\, \boldsymbol{\kappa}) = \mathcal{F}\{\tilde{t}(\boldsymbol{\rho}_{o})\},  && \tag{\ref{eq:figuer}{.1}}
\end{align*}
where $\tilde{t}(\boldsymbol{\rho}_o) = \exp\left[\frac{j\pi}{\lambda z_2} |\boldsymbol{\rho}_o|^2 + \frac{j\pi z_1}{\lambda z_2^2} |\boldsymbol{\rho}_o|^2 + \frac{j\pi}{\lambda z_3} |\boldsymbol{\rho}_o|^2 \right] \tilde{T}(\boldsymbol{\rho}_o)$ represents the effective target response, defined as the product of the target’s complex reflectivity and the quadratic phase factors arising from free-space propagation. Furthermore, $\boldsymbol{\rho}_r^{(m,n)}$ designates the discretized reference arm coordinates at index $(m,n)$, while $\boldsymbol{\kappa}_{m^{'},n^{'}}$ denotes the discretized wave vector at index $(m^{'},n^{'})$. A detailed derivation is provided in Supplement 1, Sec. S3.B. Upon retrieving the complex reflectivity distribution, the total image energy is quantified by summing the intensities across all pixels: $\sum_{m,n} |G[\boldsymbol{\rho}_r^{(m,n)}]|^2$.

As evidenced by Eqs. \eqref{eq:f_if} and \eqref{eq:figuer}, the energy of the IF signal is determined by the modulus squared of the integrated multi-mode light field: $|\widetilde{i}^{{(i)}}(f=f_{if})|^2$. In contrast, field correlation allows for the decomposition of the IF signal into individual pixels; the image energy is then calculated by taking the modulus squared of each resolved pixel and summing the results: $\sum_{m,n} |G[\boldsymbol{\rho}_r^{(m,n)}]|^2$. Crucially, while the IF signal energy of the bucket-detector-based CD-GI lidar remains constrained by the Siegman antenna theorem, the application of field correlation decouples the multi-mode fields. 

This process effectively partitions the collective IF signal into single-mode components defined by the pixel resolution, thereby circumventing the limitations in the Siegman theorem. Moreover, imaging can be achieved using a single bucket detector with a low spatiotemporal bandwidth product.

\section{Experiment} \label{experiment}
\begin{figure}[htbp]
    \centering
    \includegraphics[width=\linewidth]{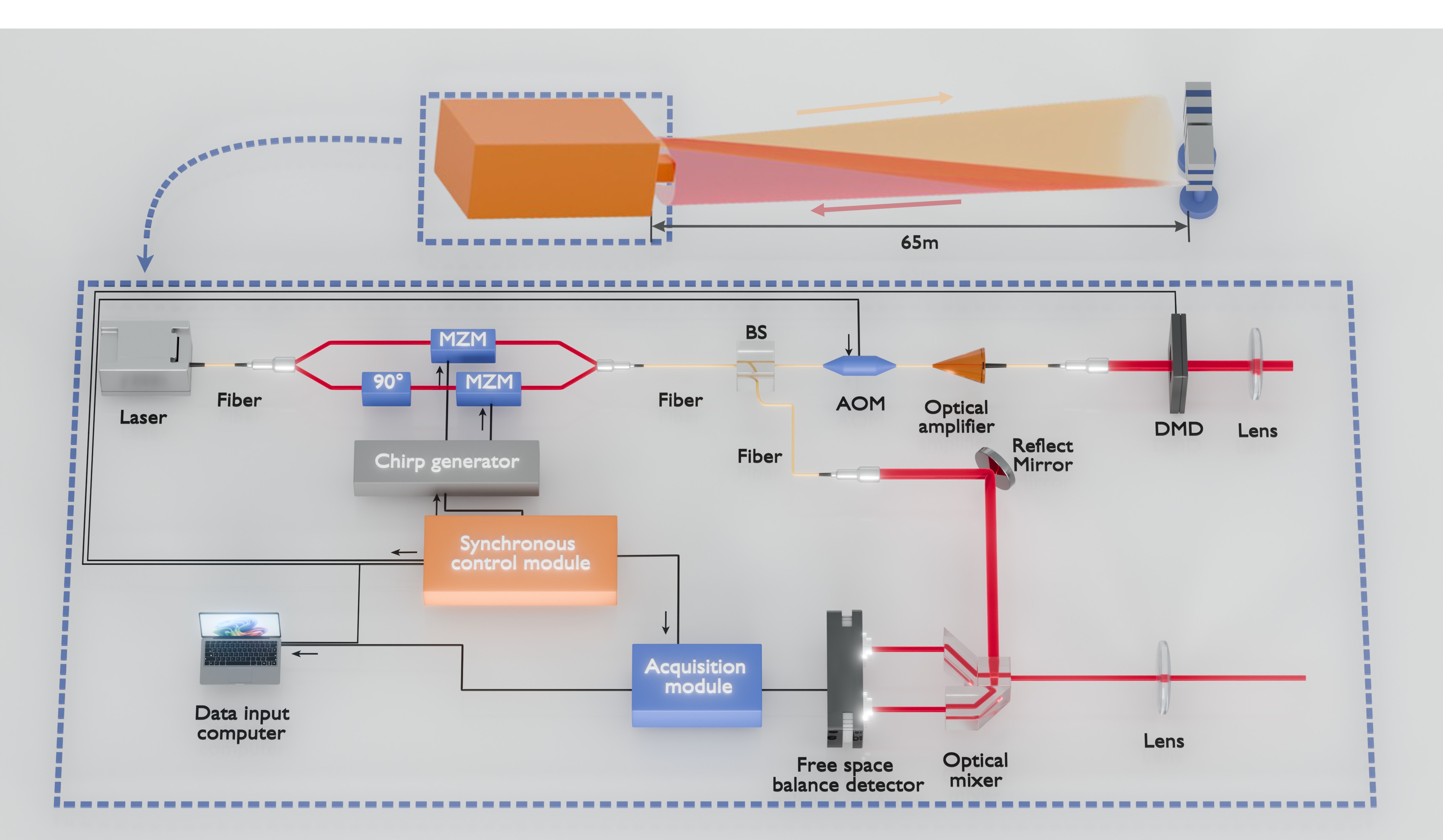}
    \caption{Experimental setup of the bucket-detector-based multi-mode CD-GI lidar system.}
    \label{fig:systems}
\end{figure}

The experimental system for the bucket-detector-based multi-mode CD-GI lidar is illustrated in Fig.~\ref{fig:systems}. A narrow-linewidth continuous-wave seed source with a central wavelength of $1550$~nm (Shanghai Precilasers Technologies Co. Ltd., FL-SF-1550-S) is coupled into a single-mode optical fiber and fed to an IQ modulator (MXIQER-LN-30, iXblue). The chirped RF signal driving the modulator is generated by a direct digital synthesizer (AD9914, Analog Devices Inc.), featuring a modulation bandwidth of $800$~MHz and a period of $1$~ms.

Following modulation, the temporally modulated light is partitioned by a fiber optic splitter (Thorlab, TN1550R1A1, 99:1) into two paths. The $1\%$ branch is collimated (Thorlab, F260APC-1550) to serve as the LO beam. The $99\%$ branch passes through an AOM (Qingjin-OE, G-1550-40-L-B-T-AA-A-Y-L) to introduce a frequency offset of $40$~MHz. The beam is subsequently amplified by an erbium-doped fiber amplifier (Shanghai Precilasers Technologies Co. Ltd.) and spatially encoded via a DMD (Texas Instruments, DLP-650-L-NIR). The modulated beam is transmitted onto the target through a simple single-lens imaging module with a focal length of 800~mm and an aperture of 150~mm. The light field reflected from the target is optically mixed with the LO beam in free space via the spatial optical mixer implemented in the imaging receiving module, which has a focal length of 250~mm and an aperture of 50.8~mm. The resulting IF signal is then obtained using a free-space balanced detector (Thorlabs, PDB230C), which the maximum photosensitive area diameter is 0.3~mm.

\subsection{Experiments of static target imaging with the bucket-detector-based multi-mode CD-GI lidar}
In the static target imaging experiments, the imaging transmission module in Fig.~\ref{fig:systems} yields a minimum resolvable feature size of approximately $1.22\lambda f/D \approx 10~\mu\mathrm{m}$. The DMD has a resolution of \(1280 \times 800\) with $10.8~\mu\mathrm{m}$ micromirrors and operates at a refresh rate of 4000~Hz. To match the resolution of imaging transmission module, the minimum unit of Hadamard pattern is set to a 1-pixel binning, resulting in an encoded block size of $10.8~\mu\mathrm{m}$. The effective coding region is $64 \times 64$ modulation array after binning. Under the experimental conditions at 65 m, the emission angle is 0.85~mrad, corresponding to the FoV of 55.5~mm.

To investigate the imaging performance of the bucket-detector-based multi-mode CD-GI lidar under different surface roughnesses of targets, experiments were conducted using both rough and smooth targets. The overall size of the targets was 55~mm. For rough target, the object consisted of the four letters “SIOM”, where the letter were fabricated using a high-reflectivity film, while the background region was composed of low-reflectivity black cardboard. For smooth target, the letter were made of black cardboard, and the background was a planar reflective mirror, forming a smooth reflective surface.
\begin{figure}[htbp]
    \centering
    \includegraphics[width=0.7\linewidth]{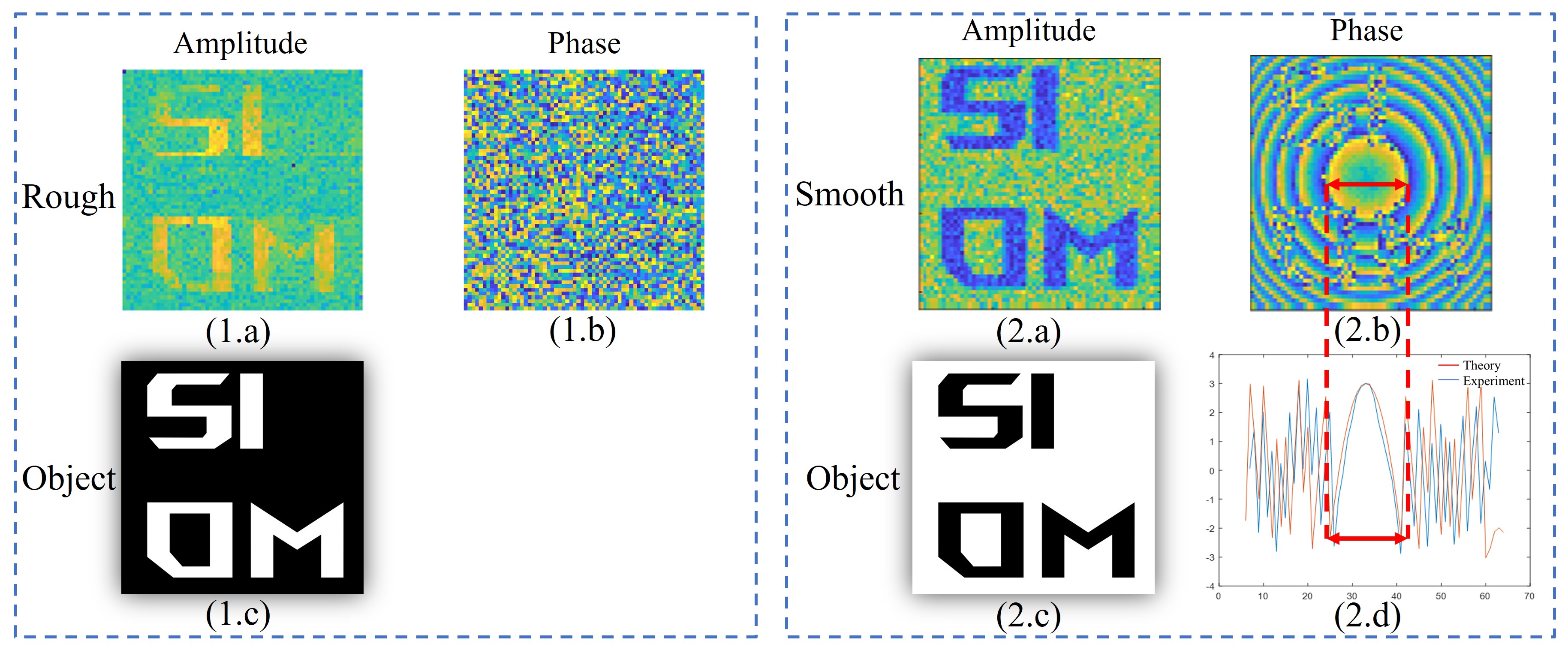}
    \caption{The reconstructed image of amplitude and phase reconstructions for rough and smooth targets. (1.a) Amplitude and (1.b) phase images of the rough target, whose ground truth object is shown in (1.c). (2.a) Amplitude and (2.b) phase images of the smooth target, with the corresponding object shown in (2.c). The smooth target exhibits a clear deterministic phase distribution, whereas the rough target produces a spatially random phase. (2.d) Comparison of the theoretical curve and experimental results of the one-dimensional phase distribution.
 }
    \label{fig:FoV_results}
\end{figure}

Fig.~\ref{fig:FoV_results} presents a comparison of the reconstructed amplitude and phase images for rough and smooth targets. For the rough target, the reconstructed amplitude image in Fig.~\ref{fig:FoV_results}(1.a) clearly reveals the “SIOM” pattern, whereas the corresponding phase image in Fig.~\ref{fig:FoV_results}(1.b) exhibits random fluctuations due to diffuse scattering from the rough surface. The ground-truth object is shown in Fig.~\ref{fig:FoV_results}(1.c) for reference.

In contrast, for the smooth target, both the amplitude image in Fig.~\ref{fig:FoV_results}(2.a) and the phase image in Fig.~\ref{fig:FoV_results}(2.b) show well-defined spatial structures. In particular, the phase image shows concentric fringe figures characteristic. To further validate the phase reconstruction, Fig.~\ref{fig:FoV_results}(2.b) and Fig.~\ref{fig:FoV_results}(2.d) show that the experimental results agree well with the term $\exp\left[ \frac{j \pi}{\lambda} \left( \frac{z_1 + z_2}{z_2^2} + \frac{1}{z_3} \right) |-\frac{z_2}{z_1}\boldsymbol{\rho}_r|^2 \right]$ in Eq.\eqref{eq:f_if}, which is caused by both the simple single-lens imaging transmission module and the imaging receiving module.

Next, we experimentally investigate the IF signal energy and the energy of the reconstructed images in the bucket-detector-based multi-mode CD-GI lidar system.

\subsection{Experiments of IF signal energy and reconstructed image energy with the bucket-detector-based multi-mode CD-GI lidar system}
As shown in Eq.~\eqref{eq:f_if}, to investigate the relationship among the imaging receiver aperture, detector area, and the IF signal energy, a series of black pinhole masks with diameters of 0.1~mm, 0.2~mm, and 0.3~mm (Edmund Optics) were sequentially placed on the detector plane to control the effective detection area. These pinhole diameters corresponded to FoV diameters of 25.9~mm, 51.8~mm, and 77.7~mm on the target plane at 65~m. The minimum unit of Hadamard pattern is set to a 4-pixels binning. The effective coding region is $64 \times 64$ modulation array after binning. To regulate the receiving aperture, a variable diaphragm is positioned before the lens within the imaging receiving module. The target was a circular reflectivity plate with a diameter of 100~mm, ensuring uniform filling of the maximum effective detection area of the detector. The IF signal energy was obtained from the same number of repeated measurements of the first fully illuminated pattern followed by statistical averaging, while the reconstructed image energy was computed by converting the reconstructed intensity into relative energy values. The pixel count of the target area is obtained by normalizing the reconstructed image to its maximum intensity and applying a 2\% intensity threshold (0.02) to classify and isolate the target pixels.

\begin{figure}[htbp]
    \centering
    \includegraphics[width=0.8\linewidth]{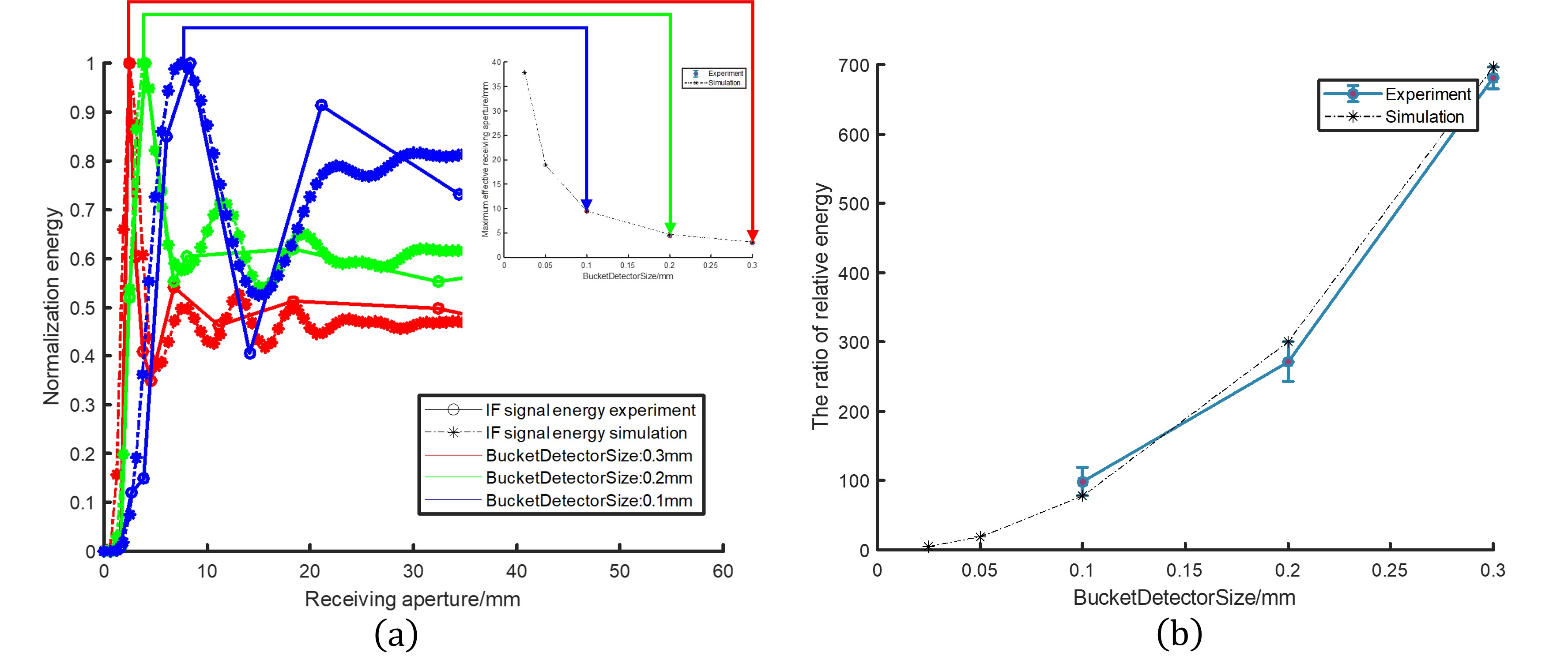}
    \caption{
    Normalized IF signal energy and the relative energy ratio of the reconstructed images versus the receiving aperture variation range under different bucket detector diameters (simulation and experiment). (a) Normalized IF signal energy as a function of the receiving aperture variation range for various bucket detector diameters: blue represents a detector diameter of 0.1~mm; green represents a detector diameter of 0.2~mm; red represents a detector diameter of 0.3~mm; the circle represents the experimental curve; the asterisk represents the simulation curve. The horizontal axis represents the receiving aperture, and the vertical axis represents the normalized energy. (b) Relative energy ratio of the reconstructed images as a function of the bucket detector diameter: the asterisk denotes the simulation curve, while the blue solid line represents the experimental curve. The horizontal axis represents the bucket detector diameter, and the vertical axis represents the ratio of relative energy.
    }
    \label{fig:guaid1}
\end{figure}

We confirmed a clear matching relationship between the receiver aperture and detector size that maximizes the IF signal energy, as described by Eqs.~\eqref{eq:sighenb} and ~\eqref{eq:sighenb1}. The experimental results are shown in Fig.~\ref{fig:guaid1} (a). Under single-mode LO conditions, the receiving apertures for detectors with diameters of 0.3~mm, 0.2~mm, and 0.1~mm were measured to be 3.00~mm, 4.50~mm, and 9.50~mm, closely matching the theoretical predictions described by Eqs.~\eqref{eq:sighenb} and ~\eqref{eq:sighenb1} of 3.15~mm, 4.73~mm, and 9.46~mm. This result is also in agreement with the effective mode field matching conditions predicted by the Siegman antenna theorem\cite{siegman1966antenna}.
Furthermore, the statistical characteristics of the IF signal energy in the bucket-detector-based multi-mode CD-GI lidar system and its relationship with the total energy of the reconstructed image by field correlation have been systematically evaluated. The experimental results in Fig.~\ref {fig:guaid1} (b) show that the ratio of the total energy of all reconstructed image pixels to the IF signal energy varies with detector size. Specifically, the measured ratios were 680.9 for $D_{\mathrm{det}} = 0.3 $~mm, 271.6 for $D_{\mathrm{det}} = 0.2$~mm and 98.5 for $D_{\mathrm{det}} = 0.1 $~mm. These ratios closely match the theoretical expectations based on the effective number of pixels shown in Eq.~\eqref{eq:figuer} (697, 300 and 78, respectively, as detailed in Supplement~1, Sec.~S4.A), directly validating the ability to separate the sum of the reflected multi-mode light fields to pixels using field correlation. 

Experiments on the IF signal energy and image energy from field-correlation reconstruction jointly demonstrate that, although the IF signal energy is constrained by Siegman antenna theorem, the exploitation of field correlation enables the reflected multi-mode light fields collected by the bucket detector to be coherently decoupled into effective single-mode components determined by the pixel size, thereby effectively circumventing the constraint imposed by Siegman antenna theorem.

\subsection{Experiments of vibration modes imaging}

The bucket-detector-based multi-mode CD-GI lidar is further applied in experiments to assess its capability in vibration modes imaging of the target.

\subsubsection{Experiments of one vibrating target:}
\begin{figure}[!htbp]
    \centering
    \includegraphics[width=0.8\linewidth]{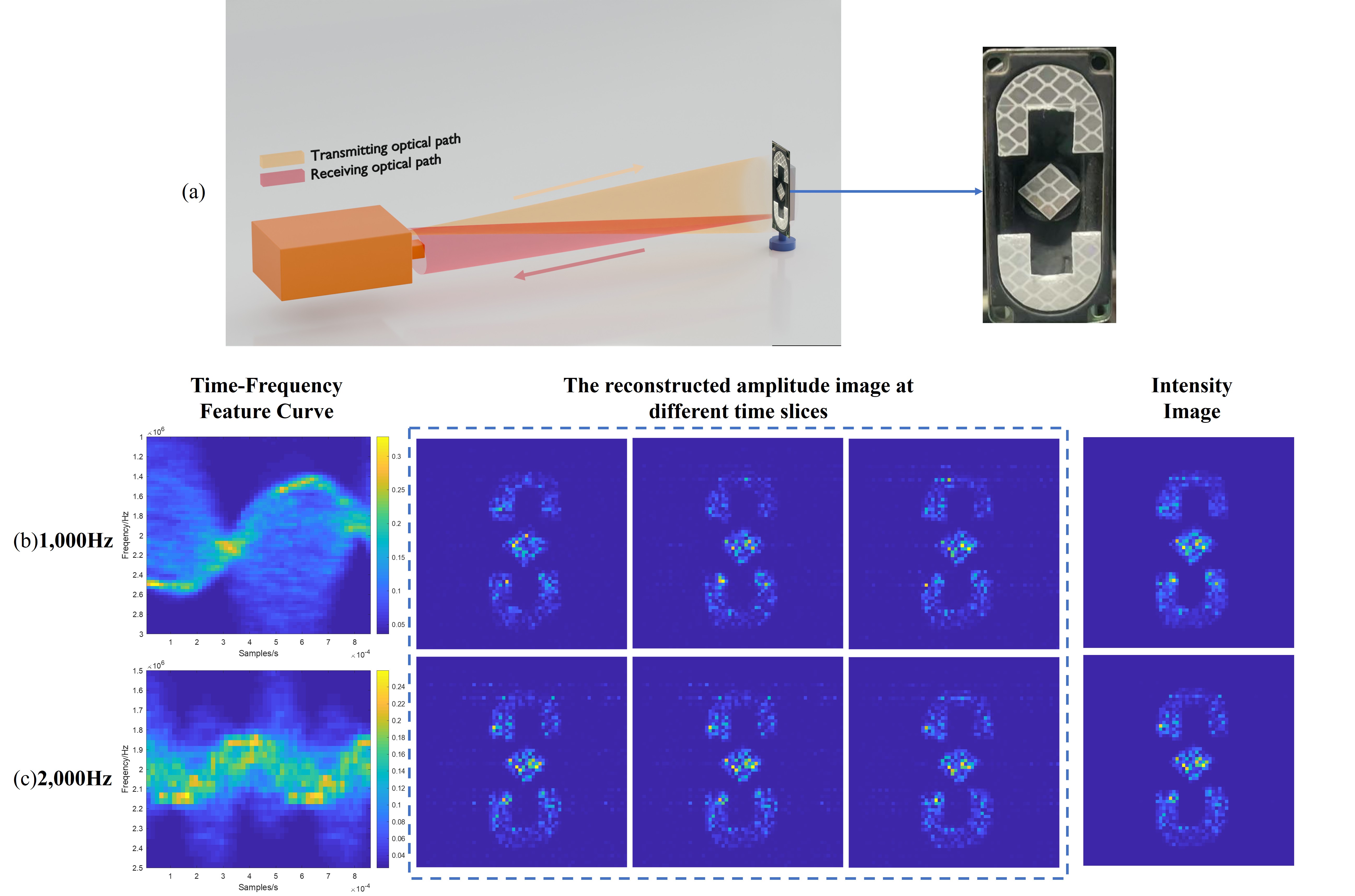}
    \caption{Experimental results of spatial vibration mode reconstruction of the vibrating target: (a) simplified schematic of one vibrating target setup; (b) vibration frequency of 1000~Hz; (c) vibration frequency of 2000~Hz.
    }
    \label{fig:object_modes}
\end{figure}

A loudspeaker (8~$\Omega$, 5~W) was employed as the vibration source. Because the diaphragm material of the speaker has very low reflectivity at 1550~nm, a high-reflectivity (HR) film was affixed to its surface to enhance the optical reflectivity. The target reflectivity distribution was set as illustrated in Fig.~\ref{fig:object_modes}(a), featuring a rectangular geometry with a length of $60$~mm and a width of $25$~mm. This specific target scale was chosen to ensure it remains well within the detection FoV at the $65$~m. The loudspeaker was driven by a sinusoidal voltage signal generated from a computer sound card, with vibration frequencies set at 1000~Hz and 2000~Hz. To capture as much as possible the entire vibration cycle and record the complete vibration state of the target, the repetition frequency of the DMD is set at $1000$~Hz. The effective coding region is $64 \times 64$ modulation array after 2-pixels binning (Hadamard patterns). The detection FoV is measured at $77$~mm. The sampling rate is established at $125$~MHz to satisfy the Nyquist sampling theorem for the $40$~MHz IF signal.

For each pattern acquisition, $110,016$ data points are collected, corresponding to a total duration of $880~\mu\text{s}$. The window length for the STFT is defined as $4096$ points (approximately $33~\mu\text{s}$). Within this localized time interval, the target motion is approximated as uniform velocity (see Supplement~1, Sec.~S2.B for further details). An overlap length of $2048$ points (approximately $17~\mu\text{s}$) is applied between adjacent windows to ensure temporal continuity.
\begin{table}[htbp]
\centering
\caption{Key Experimental and Signal Processing Parameters}
\label{tab:parameters}
\begin{tabular}{lcc}
\hline
\textbf{Parameter} & \textbf{Value} & \textbf{Unit} \\ \hline
DMD Repetition Frequency & 1000 & Hz \\
Sampling Rate & 125 & MHz \\
IF Signal Frequency & 40 & MHz \\
Points per Each Pattern & 110,016 & - \\
Total Acquisition Time & 880 & $\mu$s \\
STFT Window Length & 4096 (33) & pts ($\mu$s) \\
STFT Overlap Length & 2048 (17) & pts ($\mu$s) \\ \hline
\end{tabular}
\end{table}

The data processing workflow is implemented as follows:

\begin{enumerate}
    \item \textbf{Time-Frequency Representation:} The electrical signals corresponding to 8,192 patterns are processed via STFT using a rectangular window ($l=4096$, $d=2048$). 
    
    \item \textbf{Turbulence Compensation:} To mitigate phase jitter induced by atmospheric turbulence, the resulting STFT spectra are processed according to Eq.~\eqref{correct_signal}. Specifically, the phase fluctuations are compensated by performing point-wise multiplication between the STFT spectra of odd frames and the complex conjugate of the even frames.
    
    \item \textbf{Micro-Doppler Extraction:} The modules of the 4,096 compensated spectra are accumulated and normalized. A global thresholding algorithm is then applied to the normalized spectrogram to identify the support set $\omega_k^{(h)}$, which represents the time-frequency locations of all micro-Doppler components.

    \item \textbf{Coordinate and Height Imaging:}
    \begin{itemize}
        \item \textbf{Step 1:} The identified micro-Doppler frequencies are converted into a vibration height matrix $A_k^{(h)}$ using Eq.~\eqref{eq:ampli}.
        \item \textbf{Step 2:} A looping algorithm extracts the complex signals at $f=\frac{\omega_k^{(h)}}{2\pi}$ from each compensated STFT spectrum to serve as the bucket signals.
    \end{itemize}

    \item \textbf{Field Correlation Reconstruction:} For each temporal slice $k$, the bucket signals are correlated with the reference fields. This process reconstructs the complex reflectivity of the target at specific micro-Doppler frequencies. The spatial coordinates are determined by extracting the intensity of the reconstructed images.
    \item \textbf{Spatiotemporal Synthesis:} By integrating the vibration height matrix with the corresponding spatial coordinates across all temporal slices, the full-field spatial distribution of the target's vibration mode is successfully retrieved.
\end{enumerate}

Fig.~\ref{fig:object_modes} (b) and (c) display the vibration characteristics of the target at different excitation frequencies, including the time frequency spectrum, the reconstructed amplitude image at different time slices, and the intensity image obtained by summing the intensity images from different time slices. Since the reconstructed vibration height images are dynamic, they are provided as videos in the Supplementary Materials (Visualizations 1.1 and 1.2). These results demonstrate that the vibration mode of the target can be successfully reconstructed.

\begin{figure}[!htbp]
    \centering
    \includegraphics[width=0.8\linewidth]{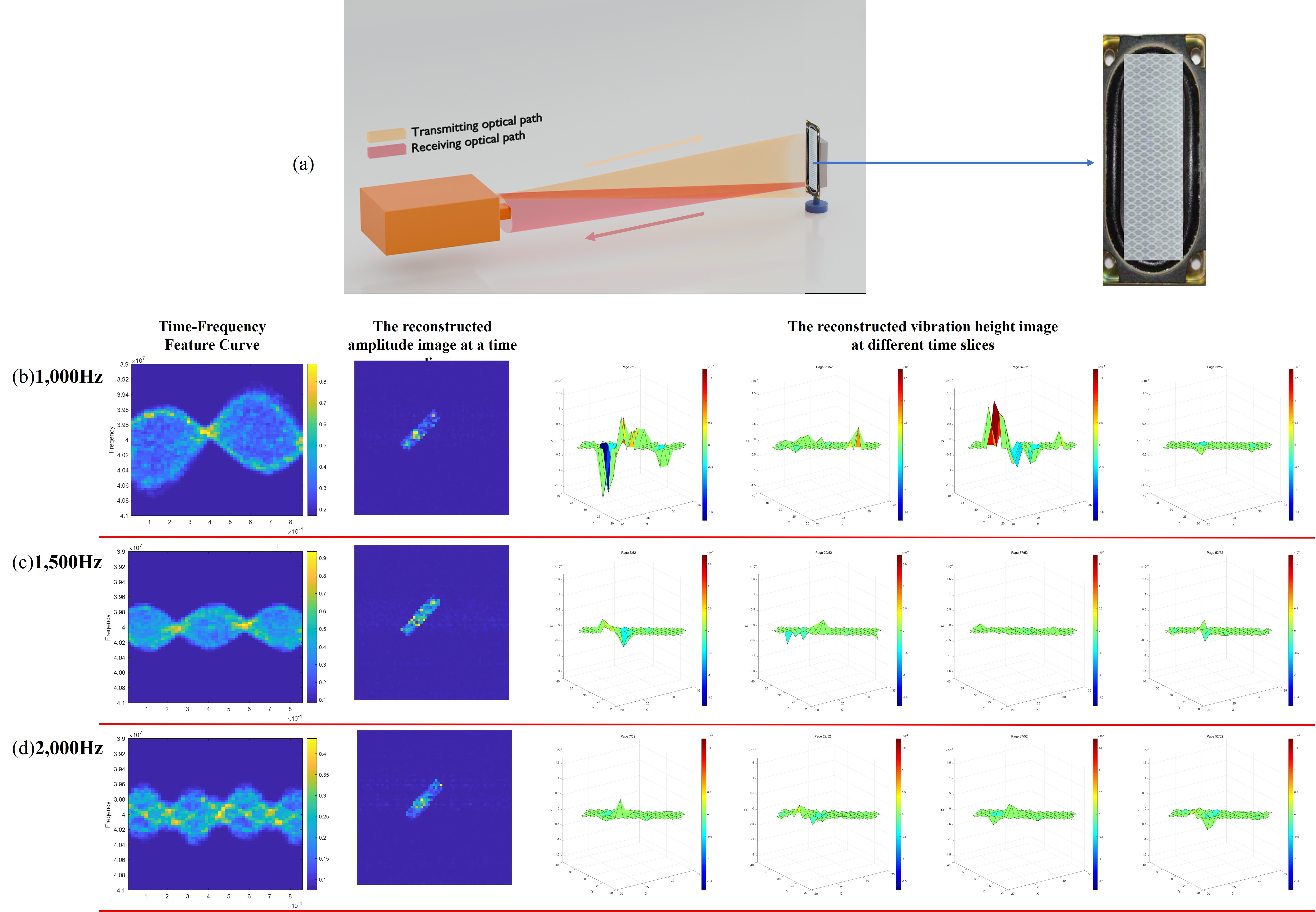}
    \caption{Simplified schematic of the single vibrating target setup: (a), the vibrating feature information fusion diagram for one vibrating target, and each row corresponds to a different vibration frequency: (b)–(d) represent 1000~Hz, 1500~Hz, and 2000~Hz, respectively, from top to bottom. First column: Time-frequency distribution profiles characterizing the target vibration. Second column: Reconstructed amplitude images of the target's complex reflectivity obtained through field correlation. Third column: Corresponding vibration height images reconstructed at different time slices.}
    \label{fig:multidimensional_modes}
\end{figure}
To investigate the changes in vibration modes at different driving frequencies, the target was configured as a strip-shaped object measuring 7~mm in width and 60~mm in length. The HR film was clamped at both ends, while the central region was adhered to a diaphragm using adhesive tape, forming a planar vibrating target. The effective coding region is $64 \times 64$ modulation array after 4-pixels binning (Hadamard patterns). The detection FoV is measured at $77$~mm. The repetition frequency of the DMD is set at $1000$~Hz. A simplified schematic of the experimental setup is shown in Fig.~\ref{fig:multidimensional_modes} (a). The vibration frequencies were set at 1000~Hz, 1500~Hz, and 2000~Hz.

According to the vibration modes imaging reconstruction approach for vibrating targets, Fig.~\ref{fig:multidimensional_modes} (b)-(d) illustrates the reconstructed vibrating feature information of the target at different vibration frequencies. Under a fixed driving power, the vibration height of target gradually decreases with increasing vibration frequency\cite{jacobsen1930steady}, which aligns with the physical behavior of a damped driven harmonic oscillator. Since the reconstructed vibration images are dynamic, the complete temporal evolutions are provided as Supplementary Media (Visualizations 2.1, 2.2, and 2.3). For static demonstration in the manuscript, representative frames are sampled from the sequences with an initial offset at the 7th frame and a step size of 15 frames.

\subsubsection{The Imaging Experiment for Distinguishing Multiple Vibration targets}

\begin{figure}[!htbp]
    \centering
    \includegraphics[width=0.9\linewidth]{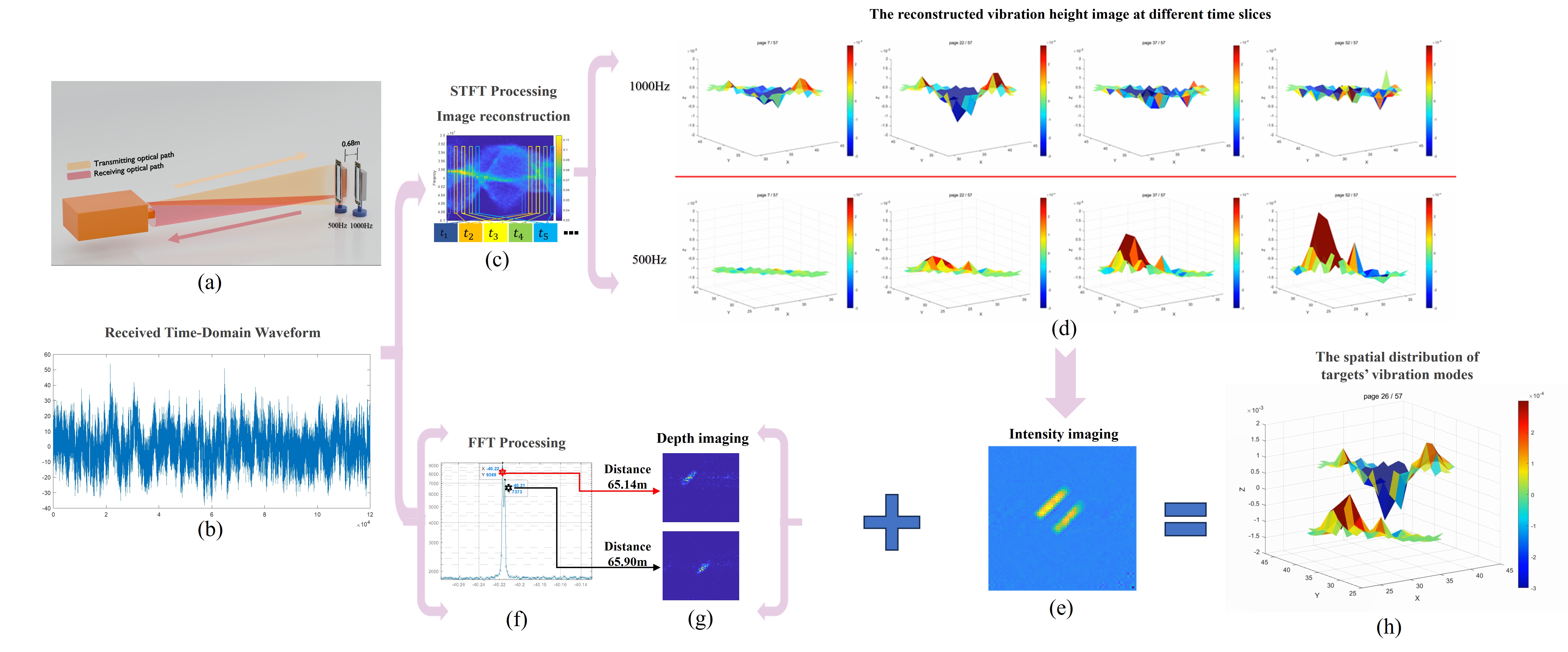}
    \caption{
    Results of the vibration frequencies, vibration mode spatial distributions, 2D spatial distributions, and distances of two targets with different vibration frequencies (For the dynamic video, see Visualization 3). (a) Schematic of the experimental setup. (b) Received time-domain waveform from the balanced bucket detector (Taking the electrical signal corresponding to a single pattern as an example). (c) STFT processing. (d) Reconstructed vibration height images at different time slices corresponding to 1000~Hz and 500~Hz targets. (e) Reconstructed intensity images of the target's complex reflectivity. (f) FFT processing for distance extraction. (g) Reconstructed depth images for two targets at 65.14~m and 65.90~m. (h) The reconstructed spatial distribution of two targets’ vibration modes.
    }
    \label{fig:double_modes_show}
\end{figure}

To further evaluate the system’s capability to simultaneously resolve multiple vibrating targets, two strip-shaped vibrating targets with identical dimensions (7~mm in width and 60~mm in length) were positioned at different axial distances and independently driven at 500~Hz and 1000~Hz, respectively. The corresponding target distances were 65.00~m and 65.68~m. The effective coding region is $64 \times 64$ modulation array after 4-pixels binning (Hadamard patterns). The repetition frequency of the DMD is set at $1000$~Hz. The detection FoV is measured at $77$~mm. A simplified schematic of the experimental setup is shown in Fig.~\ref{fig:double_modes_show} (a).

Fig.~\ref{fig:double_modes_show} presents the results of the vibration frequencies, vibration mode spatial distributions, 2D spatial distributions, and distances of two targets with different vibration frequencies. The raw time-domain waveform captured by the balanced bucket detector is presented in Fig.~\ref{fig:double_modes_show} (b). The corresponding time-frequency spectrum, obtained by performing STFT on the time-domain waveform, is shown in Fig.~\ref{fig:double_modes_show} (c). First, feature extraction is performed on this spectrum. By employing a curve-fitting approach, the vibration frequencies of the two targets are identified as 495 Hz and 1007 Hz.
Next, based on temporal slicing, the vibration height corresponding to each frequency component is calculated as shown in Fig.~\ref{fig:double_modes_show} (d), enabling the separation and reconstruction of the time-varying vibration height images of the two targets with distinct vibration frequencies. The dynamic evolution of the target vibration modes is provided in Visualization 3. Fig.~\ref{fig:double_modes_show} (e) represent the reconstructed intensity image of the targets' complex reflectivity.

Finally, pulse compression could also be applied to the electrical signal, which involves performing a FFT over the entire signal to extract the distance information of the two targets as shown in Fig.~\ref{fig:double_modes_show} (f) and (g). The reconstructed target distances were 65.14~m (500~Hz) and 65.90~m (1000~Hz), yielding a longitudinal separation of 0.76~m. Compared with the actual separation of 0.68~m, the resulting deviation of 0.08~m falls well within the system’s distance resolution of $\Delta z = c/(2B) \approx 0.187$~m, indicating reliable distance discrimination. Combining the distance information with the reconstructed vibration modes allows for more precise target characterization. Compared with intensity image, our system enables simultaneous acquisition of spatial distribution of vibration modes and target reflectivity.

\section{Conclusion and Discussion}
In summary, this work presents a multi-mode CD-GI lidar strategy based on bucket detector. It breaks the limitations of Siegman antenna theorem by exploiting field correlation to decouple the coherent summation of reflected multi-mode light fields. The results demonstrate that this strategy reconstructs the spatial distribution of target’s complex reflectivity with high sensitivity under the spatiotemporal bandwidth product of a single bucket detector. This strategy not only discriminates targets according to their vibration frequencies, but also reconstructs the spatial distribution of targets’ vibration modes. Moreover, the depth imaging and three-dimensional imaging capability is preserved, enabling a more comprehensive characterization of the target’s spatiotemporal characteristics.

By simultaneously acquiring time--frequency features and spatial vibration mode distributions, the system provides enriched target descriptors that enable more accurate recognition, classification, and comprehensive target analysis. This work shows potential in applications such as autonomous sensing, non-contact structural health monitoring, precision micro-vibration measurement, and multi-modal remote sensing.

\begin{backmatter}
\bmsection{Funding}
This work is supported by the National Key R\&D Program of China (JCKY2024110C034).

\medskip

\bmsection{Acknowledgment}
The authors are grateful to Professor Wenlin Gong for his constructive discussions.

\medskip

\bmsection{Disclosures}
The authors declare no conflicts of interest.

\medskip

\bmsection{Data Availability Statement}
Data underlying the results presented in this paper are not publicly available at this time but may be obtained from the authors upon reasonable request.

\medskip

\bmsection{Supplemental document}
See Supplement 1 and Visualization for supporting content.

\medskip

\end{backmatter}

%%%%%%%%%%%%%%%%%%%%%%% References %%%%%%%%%%%%%%%%%%%%%%%%%

%%%%%%%%%% If using BibTeX:
\bibliography{sample}

%%%%%%%%%% If preparing manually:
% \begin{thebibliography}{1}
% \newcommand{\enquote}[1]{``#1''}

% \bibitem{Zhang:14}
% Y.~Zhang, S.~Qiao, L.~Sun, Q.~W. Shi, W.~Huang, L.~Li, and Z.~Yang,
%   \enquote{Photoinduced active terahertz metamaterials with nanostructured
%   vanadium dioxide film deposited by sol-gel method,}
%   {\protect\JournalTitle{Optics Express}} \textbf{22}, 11070--11078 (2014).

% \bibitem{Optica}
% {Optica}, \enquote{{Optica Publishing Group},}
%   \url{http://www.opg.optica.org}.

% \bibitem{FORSTER2007}
% P.~Forster, V.~Ramaswamy, P.~Artaxo, T.~Bernsten, R.~Betts, D.~Fahey,
%   J.~Haywood, J.~Lean, D.~Lowe, G.~Myhre, J.~Nganga, R.~Prinn, G.~Raga,
%   M.~Schulz, and R.~V. Dorland, \enquote{Changes in atmospheric consituents and
%   in radiative forcing,} in \enquote{Climate Change 2007: The Physical Science
%   Basis. Contribution of Working Group 1 to the Fourth Assesment Report of
%   Intergovernmental Panel on Climate Change,}  S.~Solomon, D.~Qin, M.~Manning,
%   Z.~Chen, M.~Marquis, K.~B. Averyt, M.~Tignor, and H.~L. Miler, eds.
%   (Cambridge University Press, 2007).

% \end{thebibliography}

\end{document}